\documentclass[aps,prd,reprint,nofootinbib,longbibliography,showkeys]{revtex4-1}
\usepackage{blindtext}
\usepackage{mathtools}
\usepackage{xcolor}

\usepackage[utf8]{inputenc}
\usepackage[english,brazil]{babel}
\usepackage{hyperref}
\definecolor{mypink1}{rgb}{0.858, 0.188, 0.478}
\definecolor{mypink2}{RGB}{219, 48, 122}
\hypersetup{
    colorlinks=true,
    linkcolor=blue,
    filecolor=blue,      
    urlcolor=red,
    citecolor=blue,
}
\usepackage{color,colortbl}
\definecolor{LightCyan}{rgb}{0.88,1,1}
\usepackage{tensind}
\tensordelimiter{?}

\usepackage{graphicx}
\usepackage{hyperref}
\DeclareGraphicsExtensions{.bmp,.png,.jpg,.pdf}
\usepackage{amsmath}
\usepackage{courier}
\usepackage{verbatim}
\usepackage{amssymb}
\usepackage{amsfonts}
\usepackage{natbib}

\begin{document}
\title{As Astrocientistas \\
e o paradoxo da desigualdade de gênero na física}
\author{Carla R. Almeida$^{1\dagger}$, Paola C. M. Delgado$^{2\ddagger}$, Tays Miranda$^{3,4\dagger *}$}

\affiliation{$^1$Instituto de Física, Universidade de São Paulo, 05314-970, São Paulo, Brasil\\
$^2$Faculty of Physics, Astronomy and Applied Computer Science,
Jagiellonian University, 30-348 Krakow, Poland\\
$^3$ Department of Physics, P.O.Box 35 (YFL), FIN-40014 University of Jyv$\ddot{a}$skyl$\ddot{a}$, Finland\\
$^4$Helsinki Institute of Physics, P.O. Box 64, FIN-00014 University of Helsinki, Finland}

\begin{abstract}
\textbf{Resumo}\\
Em fevereiro de 2021 ocorreu a primeira edição do evento \textit{As Astrocientistas: Encontro Brasileiro de Meninas e Mulheres da Astrofísica, Cosmologia e Gravitação}, que reuniu grandes nomes da ciência brasileira para homenagear as pesquisadoras brasileiras ou estrangeiras que tenham vínculos com o nosso país, em celebração ao dia internacional das mulheres e meninas na ciência. Neste artigo, apresentamos nossa motivação para organizar este evento, assim como as dificuldades que enfrentamos e os aprendizados que conquistamos. Também expomos nossa interpretação sobre a relevância deste encontro.

\textbf{Abstract}\\
The first edition of the event \textit{As Astrocientistas}: the Brazilian Meeting of Women and Girls in Astrophysics, Cosmology, and Gravitation, happened in February 2021. It gathered great names of Brazilian science to pay homage to Brazilian female researchers and other female physicists with close ties to our community. This paper presents our motivation to organize this event, the challenges we faced, and exposes the learning experience we gained from doing so. We will talk about the relevance of this event as we perceive it.
\end{abstract}

\maketitle

\section{Introdução}
A baixa representação de mulheres na ciência é paradoxal. Os conceitos que movem a prática científica, falseabilidade e reprodutibilidade, são atividades dinâmicas que exigem a colaboração de uma comunidade especializada para acontecer. O sucesso de uma teoria depende dela ser testada e questionada de diversos ângulos, com uma variedade de pensamentos que apenas um ambiente diversificado pode oferecer. Ainda assim, continua sendo comum observarmos grupos inteiros de pessoas tendo suas contribuições apagadas ou desvalorizadas simplesmente por serem diferentes do padrão dominante.

Felizmente, a questão de gênero vem ganhando espaço para debates nas últimas décadas, apesar dos obstáculos. Pesquisas sobre a participação e atuação por gênero alcançaram maior notoriedade há menos de vinte anos e são essenciais para entender e dar dimensão ao problema. Estes são os primeiros passos em busca de soluções. Porém, tais pesquisas não são devidamente valorizadas 
por uma parcela da comunidade científica. Em muitos casos, são literalmente desencorajadas. Essa oposição é apenas um dos fatores que dificultam a coleta de dados, em especial em países com uma maior resistência em discutir questões de gênero.

No Brasil de 2020, um país assolado por uma crise política, econômica, social e da saúde, onde a ocorrência de ataques contra grupos não-dominantes e a desvalorização da ciência se tornaram mais evidentes, nós, jovens pesquisadoras brasileiras, decidimos transformar a ansiedade para com o nosso próprio futuro em algo positivo. Assim nasceu a ideia para o evento As Astrocientistas, uma celebração das talentosas cientistas brasileiras, ou com vínculos fortes com o nosso país, que pesquisam nas áreas de astrofísica, cosmologia e gravitação.

A primeira edição do evento aconteceu  em formato virtual em fevereiro de 2021, mês em que se comemora o Dia Internacional das Mulheres e Meninas na Ciência, e contou com a presença de mais de cem inscritos. Com palestras de altíssima qualidade, o evento foi aclamado por sua proposta de dar visibilidade para as cientistas brasileiras e suas pesquisas. A ideia para este encontro surgiu de forma bastante ingênua e apenas aos poucos nós, integrantes do comitê organizador, começamos a entender o potencial impacto que este evento poderia trazer para jovens pesquisadoras em busca de uma carreira científica. Aprendemos muito durante a organização. Apesar de vivenciarmos esta desigualdade de gênero na academia, entendemos melhor a dimensão do problema quando o avaliamos de forma crítica. As Astrocientistas nos deu essa oportunidade e este artigo é uma dedicatória ao aprendizado que experimentamos durante o processo de organização deste evento. 

É importante frisar, no entanto, que foge ao nosso escopo neste trabalho analisar a questão de gênero na ciência de forma profunda, uma vez que não somos especialistas na área. Ainda assim, oferecemos um panorama do problema, indicando referências aos leitores, para melhor apresentarmos as diferentes questões abordadas durante a organização do evento.

Este artigo está organizado da seguinte forma: A seção \ref{dados} apresenta o problema da desigualdade de gênero em números. Escolhemos dividir esta seção em duas partes, sendo a subseção \ref{dadosglobais} voltada para pesquisas internacionais sobre o assunto. Na subseção \ref{dados_Brasil}, apresentaremos um compilado de análises quantitativas sobre a atuação de mulheres na física no Brasil, objetivando ilustrar, através de dados, a dimensão da desigualdade de gênero que a área enfrenta em nosso país, assim como a marginalização de outras minorias. Na seção \ref{Astrocientistas}, narramos a nossa experiência com a organização do evento. Também falamos sobre o processo de seleção e convite das palestrantes principais na seção \ref{convidadas}. Trazemos o resultado de uma pesquisa feita durante o evento, com a participação dos inscritos, na seção \ref{pesquisa_evento}. Por fim, apresentamos nossas conclusões e, em anexo, o título das palestras e suas respectivas autoras ou autores.

\section{O problema em números: dados sobre a desigualdade de gênero}
\label{dados}

A sub-representação de mulheres na nossa área é evidente para qualquer pessoa que já se aventurou num departamento de física. Ainda assim, a dimensão do problema se mantém velada para a maioria. Então a questão deve ser abordada através de duas perguntas motrizes: Como medir essa desigualdade? Como reduzi-la?

Atualmente, diversos grupos de pesquisadores ao redor do mundo se dedicam à coleta de dados sobre questões como assédio, segregação, falta de representatividade e etc., para então, através de uma minuciosa análise, fornecer evidências de eficácia e impacto das medidas para sanar o problema. Vamos apresentar os resultados de algumas destas pesquisas em âmbito global e nacional, a fim de salientar as proporções da desigualdade, fornecendo um contexto para a nossa motivação para a organização do evento.

\subsection{Estudos sobre a atuação de mulheres na ciência no âmbito global}
\label{dadosglobais}

De acordo com os dados mundiais, o percentual de mulheres na academia diminui consideravelmente à medida que o nível de carreira aumenta. Para tentarmos entender o porquê desta evasão desproporcional, é interessante analisar a situação através do estudo de fenômenos conhecidos como \cite{Koch-Miramond.2002}:
\begin{itemize}
	\item \textit{o diagrama de tesoura}, que representa o fato de que mulheres são ``cortadas''  da carreira de cientistas;
	\item \textit{a tubulação com vazamento}, que refere-se à situação onde as mulheres desaparecem em números desproporcionais em cada estágio da evolução de carreira acadêmica;
	\item \textit{exclusão horizontal}, que expressa a distribuição desigual de mulheres e homens em vários campos científicos;
	\item \textit{exclusão vertical}, que explicita a falta de representatividade feminina em todas os níveis de hierarquias científicas;
	\item \textit{o teto de vidro}, que expõe o desequilíbrio de gênero nas escolhas feitas pela academia.	
\end{itemize}

\begin{figure}[t]
	\includegraphics[width=\linewidth]{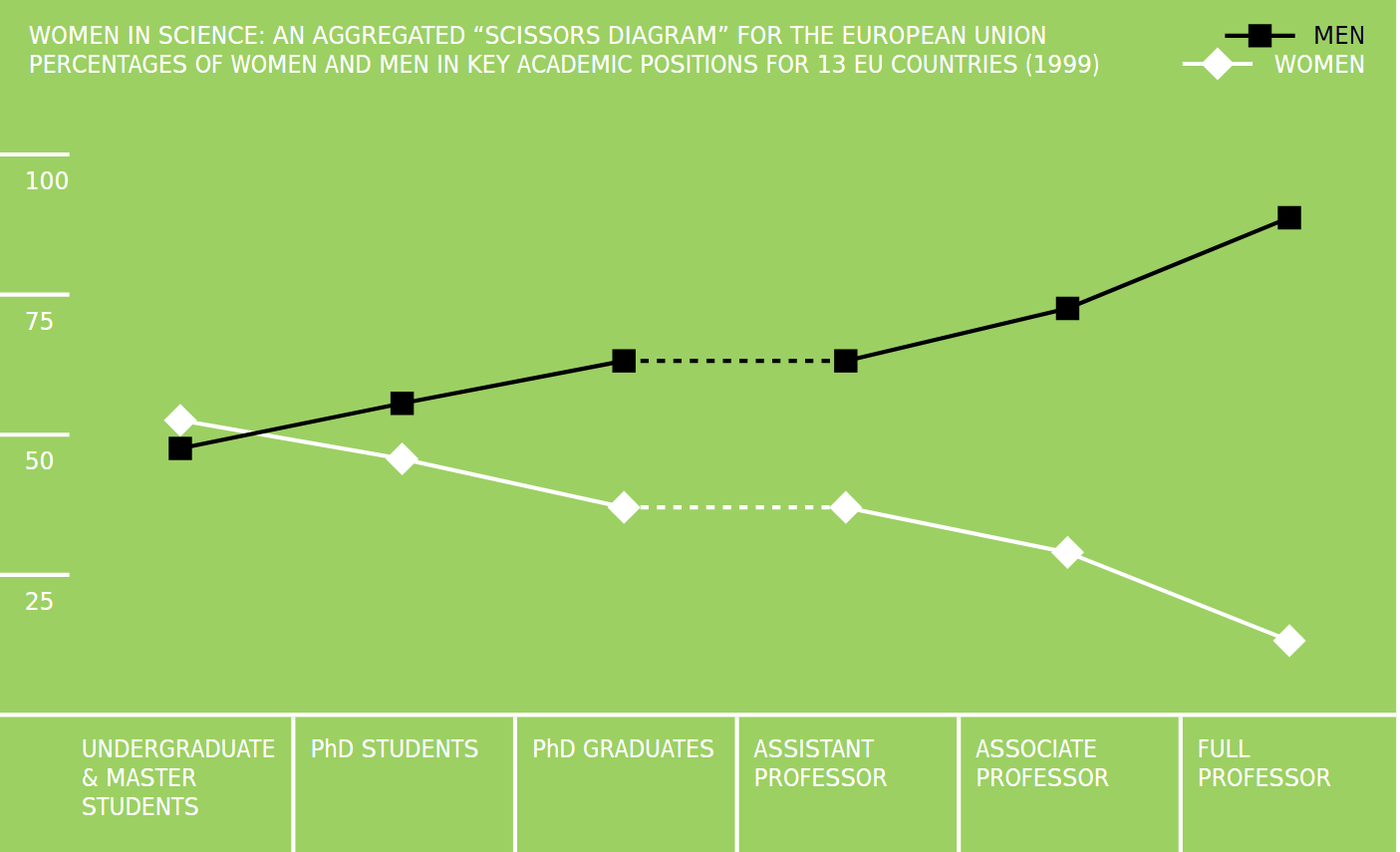}
	\caption{Em branco, o número percentual de mulheres em posições acadêmicas. Em preto, o número de homens que ocupam a mesma posição. Créditos: figura retirada da Ref. \cite{Koch-Miramond.2002}. }
	\label{tesoura}
\end{figure}

A Figura \ref{tesoura} é uma ilustração do diagrama de tesoura, evidenciando a grande diferença na percentagem de mulheres e homens em posições acadêmicas chave para $13$ países da União Europeia no ano de $1991$.

Uma das iniciativas que se propôs a identificar esses fenômenos e produzir ações para remediá-los foi a União Internacional de Física Pura e Aplicada (IUPAP, na sigla em inglês),\footnote{\href{https://iupap.org/who-we-are/}{https://iupap.org/who-we-are/}.} cujo um dos objetivos é aumentar a diversidade e inclusão na física, incentivando a participação e reconhecimento de pessoas de grupos minoritários. Desde então uma série de conferências, assembleias e resoluções vêm sendo desenvolvidas para abordar a questão.\footnote{Para maiores informações sobre as conferências de $1999-2013$, ver Ref. \cite{SBF.2015}.} A IUPAP é a única organização internacional de física que é organizada e administrada pela própria comunidade da física. Ela foi estabelecida em 1922 em Bruxelas, com 13 países membros. A primeira Assembleia Geral foi realizada em 1923 em Paris. Atualmente conta com 60 países membros, e a 29ª (e mais recente) Assembleia Geral da IUPAP foi realizada em São Paulo, Brasil, em 2017.

Na América Latina, um dos eventos visionários é o CIÊNCIA-MULHER, um evento que agregou mulheres pesquisadoras das ciências exatas e da vida para discutir a posição e participação da mulher dentro destas comunidades. A primeira conferência foi realizada no Rio de Janeiro em 2004,\footnote{Confira o site do evento no link \href{http://www.cbpf.br/~mulher/}{http://www.cbpf.br/ $\sim$mulher/}.} passando por diversos países, como, por exemplo, México, Bolívia e Guatemala.

\begin{figure}[t]
	\includegraphics[width=\linewidth]{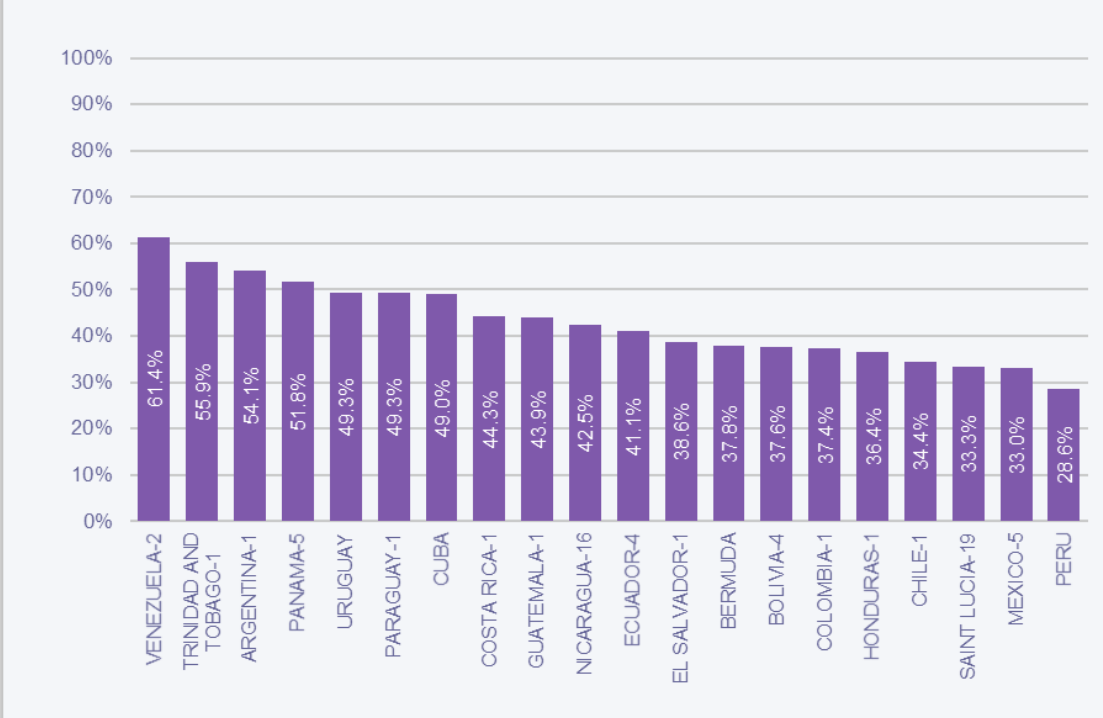}
	\caption{Percentagem de pesquisadoras mulheres com relação ao número total de pesquisadores em países da América Latina e Caribe \cite{UIS.2020}. Os indicadores $-n$ em frente ao nome dos países representam o ano em que os dados foram avaliados: ano de $(2018-n)$. Créditos: cortesia da Instituição de Estatística da UNESCO (UIS, na sigla em inglês). Fonte: \href{unesco.org}{unesco.org}.}
	\label{america}
\end{figure}

Em números, dados estatísticos sobre mulheres e ciências em 30 países europeus foram publicados pela primeira vez em $2002$ pelo Grupo de Helsinque sobre Mulheres e Ciência. No seu relatório \cite{Rees.2002}, eles fornecem informações sobre as políticas nacionais sobre as mulheres e a ciência em 15 países membros da União Europeia (UE) e nos 15 países associados ao \textit{Fifth Framework Programme} ($1996-2001$). O Instituto de Estatística da UNESCO também provê dados mundiais importantes, com pesquisas sobre a participação das mulheres na ciência na América Latina, Caribe, Europa, África, Ásia e Pacífico. De acordo com os dados de $2018$, as maiores e menores percentagens de pesquisadoras em cada região são:
\begin{itemize}
	\item Europa: a maior percentagem está na Macedônia do Norte, com 53,4\%, e a menor percentagem é encontrada na Holanda, com 26,4\%;
	\item África: a maior percentagem está na Tunísia, com 56,1\%, e a menor percentagem é encontrada no Chade, com 0,4\%;
	\item Ásia: a maior percentagem está no Myanmar, com 75,6\%, e a menor percentagem é encontrada no Nepal, com 7,8\%;
	\item Pacífico: a maior percentagem está na Nova Zelândia, com 52\%, e a menor percentagem é encontrada na Papua Nova Guiné, com 33,2\%;
\end{itemize}
A Figura \ref{america} ilustra a quantidade de mulheres pesquisadoras como uma porcentagem do número total de cientistas na América Latina e Caribe. Esses números incluem pesquisadoras em tempo parcial e em tempo integral.

É importante ressaltar que, apesar da pesquisa informar a proporção de pesquisadoras mulheres no mundo, ela não captura fatores importantes, tais quais salários, anos de emprego e lugar na hierarquia acadêmica.

Adicionalmente, a relação entre desigualdades de gênero e o departamento de física foi explorada por Cathrine Hasse e Stine Trentemøller \cite{Hasse.2011} em 2011. Os autores concluíram que diferentes culturas do local de trabalho incluem e excluem membros de acordo com a forma como consideram medidas de reconhecimento de atos criativos, tomada de risco, ``ciência útil'' e competição. Eles apontam que, não somente a cultura da sociedade é  um fator decisivo para a questão de desigualdade de gênero, mas a comunidade física em si possui hábitos que contribuem para este problema. Esta pesquisa foi realizada em cinco países europeus, Dinamarca, Estônia, Finlândia, Itália e Polônia, e três tipos de cultura foram identificados.
\begin{itemize}
	\item  Cultura de Hércules: 
	\\
	Esta é a cultura do lutador, em que o processo acadêmico é visto como extremamente competitivo e os pesquisadores têm que provar que estão no topo das suas áreas para exercer uma dada função;
	\item Cultura do cuidador:
	\\
	Esta é a cultura social, na qual o trabalho em equipe  é valorizado como fundamental para melhores resultados;
	\item Cultura de trabalho das abelhas:
	\\
	Esta é a cultura da indústria, em que pesquisadores se comportam como abelhas operárias sem muita personalidade.
\end{itemize}

Os autores encontraram uma maior proporção de mulheres físicas na Itália, onde elas constituem $33\%$ dos professores associados e $23\%$ dos professores catedráticos, e a menor proporção de mulheres físicas na Dinamarca, que constituem $10\%$ entre os professores associados e apenas $3\%$ dos professores catedráticos. A Tabela \ref{tab:1} sumariza o tipo de cultura que define os departamentos de física nestes países. Apenas a Finlândia não apresentou uma cultura definida.

\begin{table*}[t]
	\renewcommand{\arraystretch}{1.25}
	\large
	\centering
	\caption{As três culturas da física, de acordo com a Ref. \cite{Hasse.2011}.}
	\label{tab:1}
	\begin{tabular}{ c|c|c }
		\hline\hline
		Culturas &Países & $\%$ de professoras\\
		\hline
		Hércules &  Dinamarca & $3\,\%$\\
		Cuidador &  Itália & $33\,\%$\\
		Operário &  Estônia e Polônia & $11\,\%$ e $14\,\%$\\
		\hline\hline
	\end{tabular}
\end{table*}

Outro resultado interessante foi publicado por Cimpian, Kim e McDermott em 2020 \cite{Cimpian2020UnderstandingPG}, no qual eles mostram um número surpreendentemente grande de homens de baixo desempenho se formando nas disciplinas de física, engenharia e ciência da computação (PECs), comparado à participação das mulheres, nos EUA. O estudo sugere que a cultura acadêmica nessas áreas dificulta a entrada de mulheres com desempenho mediano, mas não a de homens, mostrando a irracionalidade do medo de que o aumento no número de mulheres implicaria na diminuição da qualidade na pesquisa.

Os esforços para coletar mais dados e entender melhor o problema não param. Recentemente, por exemplo, o projeto Lacuna de Gênero na Ciência --- Uma Abordagem Global para a Lacuna de Gênero em Matemática, Computação e Ciências Naturais\footnote{Original em inglês: \textit{Gender Gap in Science --- A Global Approach to the Gender Gap in Mathematical, Computing, and Natural Sciences: How to Measure It, How to Reduce It?}(\href{https://gender-gap-in-science.org/project/}{https://gender-gap-in-science.org/project/}).} colhe informações, através de pesquisas globais de cientistas, análise de publicações acadêmicas e banco de dados de boas práticas, para fornecer evidências em apoio à tomada de decisões informadas sobre políticas científicas. Outra iniciativa interessante é o \textit{website} $1400$ Degrees, lançado no dia $7$ de junho de 2021. Ele é um diretório que destaca as realizações e contribuições de mulheres e minorias marginalizadas para os campos da física e da astronomia. O nome representa a temperatura na qual alguns tipos de vidro começam a se transformar e serve como uma metáfora para a missão do projeto --- o ponto de fusão em que o teto de vidro da física se transforma em uma janela de oportunidade. Para se cadastrar na plataforma e fazer parte dessa rede de suporte, mulheres e minorias só precisam de um e-mail institucional.\footnote{O cadastro é feito em  \href{https://1400degrees.org/}{https://1400degrees.org/}.}

Em geral, estas pesquisas mundiais têm um papel fundamental de descortinar a questão de gênero e mobilizar cientistas a exigir democratização na ciência. Através de dados, relatos e divulgações de estudos, aos poucos se torna claro que, ao invés de tentar corrigir as mulheres, a comunidade acadêmica deve repensar como descrevemos e aplicamos a física. A diversidade de saberes e práticas só traz vantagens para a comunidade científica. 

\subsection{Números sobre a atuação de mulheres na física no âmbito nacional}
\label{dados_Brasil}

No Brasil, as pesquisas sobre a questão de gênero na física 
começaram a ganhar mais atração com o surgimento da Comissão de Relações de Gênero da Sociedade Brasileira de Física em 2003.
Em 2015, 
o grupo lançou o livro
{Mulheres na Física, Casos Históricos, Panorama e Perspectivas} \cite{SBF.2015}, que aborda grandes contribuições de mulheres à física, tanto no âmbito internacional quanto nacional, e apresenta dados que tornam explícita a existência de problemas a serem enfrentados pelas profissionais e estudantes da área.

Ao abordar o percentual de mulheres e homens bolsistas em programas de iniciação científica (Programa de Bolsas de Iniciação Científica - IC, e Programa Institucional de Bolsas de Iniciação Científica - PIBIC), de mestrado, de doutorado e de produtividade em pesquisa no período de 2001 a 2012, 
evidencia-se a menor participação feminina\footnote{A maior porcentagem de bolsistas mulheres é vista na categoria de bolsas de iniciação científica PIBIC em 2012, constituindo cerca de $40\%$ do total de bolsas concedidas.} e a evasão das mulheres à medida que se avança na carreira. 
No ano de 2012, as bolsas de iniciação científica PIBIC e IC concedidas a homens constituíram cerca de $64\%$ e $74\%$ do total de bolsas, respectivamente. No mesmo ano, as bolsas de mestrado e doutorado concedidas a homens representaram cerca de $80\%$ e $83\%$ do total de bolsas nas respectivas categorias Já as bolsas de produtividade (destinadas a professores) concedidas a homens no mesmo ano constituíram cerca de $88\%$ do total de bolsas. Representações gráficas com porcentagens mais precisas e informações sobre os demais anos podem ser encontradas em \cite{SBF.2015}. A desigualdade se mostra estável ao longo do período analisado, com poucas variações nos números, indicando que avanços efetivos na atuação feminina na área não têm sido alcançados.\footnote{Um sutil aumento da participação feminina é observado nas bolsas PIBIC. Entretanto, o aumento não é propagado para as bolsas de mestrado.}

\begin{figure}[h!]
	\includegraphics[width=7cm]{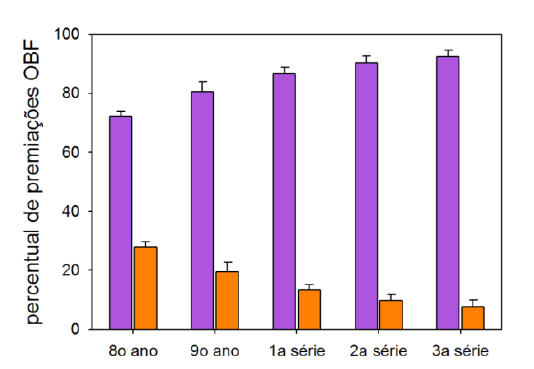}
	\caption{\label{faseescolar} Figura apresentada na Ref. \cite{Menezes.2017}. Legenda original: ``As colunas representam o percentual de mulheres (laranja) e homens (lilás) com premiações em cada ano escolar. Os valores apresentados correspondem à média calculada sobre os anos 2006-2015 (exceto de 2012-2015 para o 8º ano) e as barras de erro mostram o desvio padrão.''}
\end{figure}

O fenômeno da evasão feminina à medida que se avança na carreira ou nos estudos, conhecido como efeito tesoura ou corte vertical, é observado desde a fase escolar. Em \cite{Menezes.2017}, o número de premiações nas Olimpíadas Brasileiras de Física de 2006 a 2015 é utilizado como um indicativo da quantidade de meninas interessadas por física do 8º ano do ensino fundamental ao 3º ano do ensino médio. A média calculada sobre este período (exceto de 2012 a 2015 para o 8º ano) mostra que a porcentagem de premiações nas Olimpíadas, incluindo menções honrosas, medalhas de ouro, de prata e de bronze, é maior para os meninos em todos os anos escolares analisados, como mostra a Figura \ref{faseescolar}.

No 8º ano do ensino fundamental, cerca de $71\%$ das premiações foram concedidas a meninos. No 9º ano do ensino fundamental e no 1º, 2º e 3º anos do ensino médio, tal porcentagem sobe para cerca de $80\%$, $88\%$, $90\%$ e $91\%$. Dessa forma, os dados indicam que o corte vertical se inicia antes mesmo da escolha profissional. 

\begin{figure}
	\begin{center}
		\includegraphics[width=7cm]{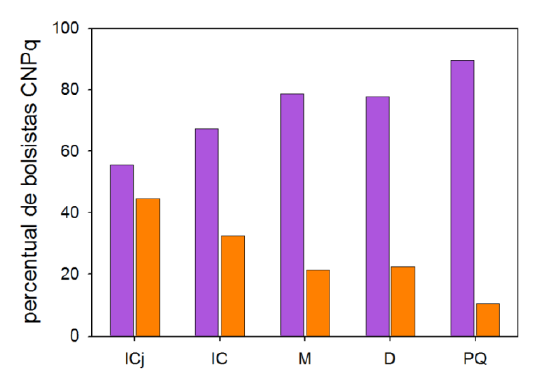}
	\end{center}
	\caption{\label{icatepq} Figura apresentada em \cite{Menezes.2017}. Legenda original: ``Percentual de bolsistas do CNPq mulheres (laranja) e
		homens (lilás): iniciação científica júnior (ICj), iniciação científica (IC), mestrado (M), doutorado (D) e
		produtividade em pesquisa (PQ).''}
\end{figure}

Em \cite{Menezes.2017} são também apresentados dados similares aos de \cite{SBF.2015} com relação a bolsas de iniciação científica, mestrado, doutorado e de produtividade em pesquisa, e dados referentes à iniciação científica júnior. Estes estão representados na Figura \ref{icatepq}.

Tal efeito tesoura pode ser uma consequência de diversos fatores. Um exemplo frequentemente mencionado 
é a maternidade. A mesma pode levar a uma diminuição da produtividade da pesquisadora ou estudante e até mesmo à desistência gerada pela sobrecarga. Essa é uma dificuldade objetiva que certamente precisa ser abordada. Entretanto, este fator 
não é único. Outros fatores sistêmicos, inclusive de caráter subjetivo, estão presentes ao longo da jornada acadêmica de forma contínua. 
Precisamos nos questionar sobre diferentes aspectos: As meninas na fase escolar têm seu interesse por física estimulado? É  dado a elas incentivo para seguir uma carreira científica? As alunas de graduação se sentem acolhidas em um ambiente muitas vezes masculinizado? Os processos seletivos de pós-graduação possuem uma proporção equilibrada de homens e mulheres aprovados? Os departamentos estão criando estratégias para aumentar o número de mulheres entre os docentes e pesquisadores? As estudantes e pesquisadoras têm espaço em reuniões de grupo, colaborações, seminários, discussões e conferências? É necessário levantar essas e outras questões, entender as dificuldades de forma clara e instituir medidas concretas e diversificadas para enfrentá-las.

Por fim, tendo em vista que a luta por igualdade e inclusão só é efetiva se englobar todas as minorias, é de fundamental importância que seja dada atenção ao espaço de grupos como o de mulheres e homens negros, homossexuais, transexuais, dentre outros, no ambiente acadêmico. Uma pesquisa realizada no período de 3 de julho a 21 de setembro de 2018, respondida por membros da Sociedade Brasileira de Física, traça um perfil da comunidade de estudantes e pesquisadores em física no Brasil através da análise de aspectos geográficos, raça, etnia, sexo, gênero, orientação sexual e deficiências \cite{Anteneodo.2020}.
\begin{figure}
	\begin{center}
		\includegraphics[width=7cm]{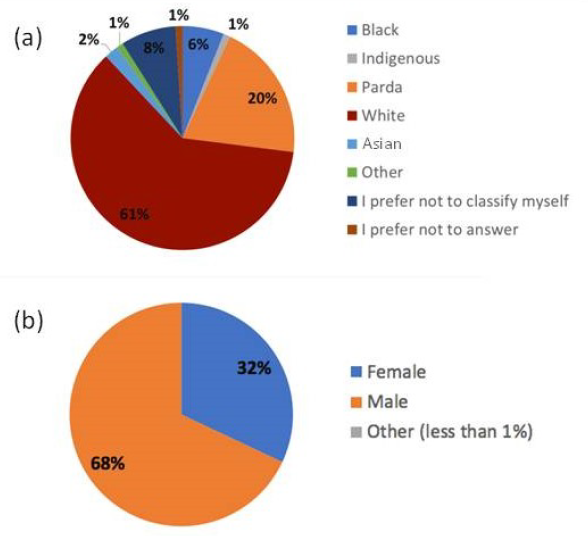}
		\includegraphics[width=7cm]{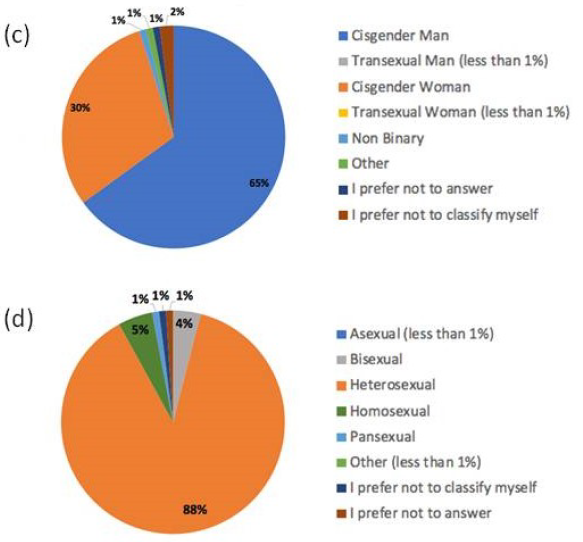}
	\end{center}
	\caption{\label{comunidadefisicabr} Figura apresentada em \cite{Anteneodo.2020}. Legenda original (tradução nossa): ``Perfil dos respondentes de acordo com etnia/raça (a), sexo (b), gênero (c) e orientação sexual (d).''}
\end{figure}

Dentre os participantes, $32\%$ responderam ser do sexo feminino. Com relação ao gênero, $30\%$ são mulheres cisgênero, menos de $1\%$ mulheres transexuais, menos de $1\%$ homens transexuais e $1\%$ não binários. No que diz respeito à orientação sexual, $5 \%$ são homossexuais, $4 \%$ bissexuais, menos de $1\%$ assexuais e $1\%$ pansexuais. Já as respostas com relação à etnia apresentam $20\%$ de pardos, $6\%$ de negros, $1\%$ de indígenas e $2\%$ de asiáticos. As opções ``Outro'', ``Eu prefiro não responder'' e ``Eu prefiro não me classificar'' também foram consideradas. Mais detalhes são apresentados na Figura \ref{comunidadefisicabr}.

Os dados mostram de forma clara o perfil da comunidade em questão: homens cisgênero ($65\%$), heterossexuais ($88\%$) e brancos ($61\%$). Na contramão, as dificuldades se mostram bastante limitadoras para os indivíduos pertencentes a uma ou mais minorias marginalizadas. Os demais resultados da pesquisa podem ser encontrados em \cite{Anteneodo.2020}.

\section{Sobre o evento As Astrocientistas}
\label{Astrocientistas}

A comissão organizadora do evento As Astrocientistas é formada pelos doutores Emmanuel Frion e Rodrigo von Marttens, além das autoras deste artigo, Carla Rodrigues Almeida, Paola Delgado e Tays Miranda. Todos nós, em um momento ou outro, fomos doutorandos em física na Universidade Federal do Espírito Santo, no Brasil. Além do vínculo institucional, somos também pesquisadores que trabalham em diversos tópicos da astrofísica, cosmologia e gravitação.

A ideia do projeto As Astrocientistas nasceu em meados de outubro de 2020, quando decidimos transformar nossa frustração com o recente retrocesso acadêmico no Brasil, impulsionado por uma crise política e a pandemia do COVID-19, em algo positivo.  Em meio a uma onda de agenda anticientífica em nosso país e a percepção do crescimento da discriminação contra grupos minoritários, parecia apropriado celebrar a ciência brasileira e as pesquisadoras que trabalham em nosso campo com uma conferência para comemorar o Dia Internacional das Meninas e Mulheres na Ciência, 11 de fevereiro. A primeira edição do As Astroscientistas -- um neologismo referente às pesquisadoras que trabalham com astrofísica, cosmologia e gravitação -- foi virtual, realizada majoritariamente em português e ocorreu entre os dias 8 e 11 de fevereiro de 2021. As Astrocientistas: Encontro Brasileiro de Meninas e Mulheres na Astrofísica, Cosmologia e Gravitação é a nossa homenagem às pesquisadoras historicamente negligenciadas que são brasileiras ou possuem vínculos com nosso país.

Nós, do Comitê Organizador, somos um grupo de pesquisadores em início de carreira trabalhando em diferentes partes do mundo e, portanto, sem uma instituição anfitriã para patrocinar este projeto. Para validar nossa causa e reafirmar às palestrantes convidadas a sinceridade do nosso convite, contamos com o apoio do Programa de Pós-Graduação em Astrofísica, Cosmologia e Gravitação da Universidade Federal do Espírito Santo (PPGCosmo - UFES), do Centro Brasileiro de Pesquisas Físicas (CBPF) e das instituições internacionais Instituto para Estudos Avançados em Ciências Humanas (KWI),\footnote{\textit{Kulturwissenschaftliches Institut Essen.}} da Alemanha, e Universidade de Jyväskylä (JyU), da Finlândia. 

O projeto As Astrocientistas começou como uma ideia muito simples que evoluiu e cresceu além da nossa imaginação. Olhando em retrospectiva, o peso político desse evento tornou-se evidente para nós quando estávamos relembrando o nosso método organizacional. Para as três mulheres do comitê organizador, a luta para usar sempre $110\%$ dos esforços para atingir os nossos objetivos em um ambiente dominado pelos homens tornou-se rotina, mas foi uma experiência nova para os dois organizadores homens, que relataram sentir uma pressão maior para obtermos um resultado de excelência. Percebemos a posteriori que, se tivéssemos falhado, não seria apenas um fracasso pessoal para nós. A nossa falta de êxito poderia enviar uma mensagem errada para nossa comunidade, podendo ser inadequadamente adotada como exemplo contra futuras tentativas de promoção de grupos minoritários.

Hoje percebemos que, apesar de todos os problemas que a pandemia gerou, a trivialização de encontros completamente virtuais nos permitiu configurar este evento de forma barata e acessível, sem a preocupação de custos com viagens e hospedagens e alcançando uma maior audiência, incluindo pessoas de fora da academia que demonstraram interesse pelo tema. Mesmo com o planejamento cuidadoso, não poderíamos ter previsto esse nível de sucesso. O entusiasmo dos participantes e as palestras de alta qualidade ministradas por nossas convidadas foram avassaladores, aumentando a pilha de evidências de que as mulheres podem fazer ciências tão bem quanto seus colegas homens, se tiverem oportunidade. 

Uma de nossas tarefas mais difíceis foi escolher quem convidar dentre tantas pesquisadores talentosas. Neste quesito, priorizamos a diversidade, entendendo que não haveria perda de qualidade seguindo este critério. Nosso objetivo foi trazer o tema da conferência de forma sensata, mas sem descuidar de sua relevância política. Com isso em mente, incluímos discussões em mesa redonda no programa para dar voz às mulheres cientistas em importantes questões científicas e políticas, inspirando a próxima geração de meninas e mulheres em astrofísica, cosmologia e gravitação. No dia final do evento, precisamente no Dia Internacional de Meninas e Mulheres na Ciência, demos as boas-vindas à ganhadora do Prêmio Nobel em física do ano de $2020$, Andrea Ghez, para uma palestra especial. Foi uma imensa honra ouvir uma das quatro cientistas que já receberam essa honraria para falar em nossa conferência. Ela é uma inspiração para toda uma geração de meninas e mulheres na física, e não podemos agradecê-la o suficiente por sua participação. 

Outro ponto de discussão durante a organização foi como abordar a participação dos homens neste evento. As Astrocientistas seria uma celebração de mulheres e meninas na ciência e entendemos isso como uma causa que pode e deve ser comemorada pelos homens também. Abrimos as inscrições para apresentações orais e pôsteres a todos, mas, para nos mantermos fiéis ao tema, pedimos aos homens que apresentassem o trabalho de mulheres cientistas, sejam estas colaboradoras ou outras pesquisadoras que eles admiram. Essas informações estavam explícitas em nosso site e nas nossas redes sociais e, ainda assim, a pergunta mais frequente que recebíamos era se homens também poderiam participar da conferência. Isso nos indicou que a comunicação não foi suficientemente efetiva, pois mesmo os interessados em homenagear as nossas convidadas não tinham a certeza do nosso objetivo, embora não saibamos o que poderíamos ter feito melhor neste sentido.

Desde o início, o registro de participantes do sexo feminino superou o número de participantes do sexo masculino por uma margem significativa. Para nós, foi a primeira vez que participamos de uma conferência de física com mais participantes do sexo feminino. Foi uma vitória! Com oportunidades e um espaço seguro, as mulheres podem prosperar neste ambiente. Mas a vitória foi agridoce ao mesmo tempo. Essa razão de mulheres para homens não reflete o cenário atual das astrociências brasileiras, que ainda é predominantemente masculina. Dada a qualidade das palestras, por que houve tão pouco interesse dos pesquisadores do sexo masculino? Esta questão permanece em aberto.

Infelizmente não fomos capazes de conseguir financiamento nacional para a produção dos Anais do evento, o que é um claro indicativo da situação precária que a ciência brasileira se encontra. A primeira edição do Astrocientistas, um evento bem sucedido que exalta pesquisadoras brasileiras de excelência, foi completamente financiado por instituições estrangeiras. Em particular, os Anais serão patrocinados inteiramente pela Universidade de Jyväskylä.

Hoje, quando o Brasil é ameaçado pelo negacionismo e os abissais cortes de verbas destinadas a ciência e tecnologia, o sucesso do Astrocientistas nos oferece um raio de esperança. Ainda há muito a aprender e melhorar em nossa sociedade, mas esses pequenos passos se mostram como a melhor maneira de seguir em frente. Esta será a primeira de muitas edições que visam criar diversidade, equidade e inclusão para a comunidade científica brasileira.

\section{Convidadas}
\label{convidadas}

A fim de homenagear e dar visibilidade às pesquisadoras brasileiras ou com vínculos fortes com o Brasil, o comitê organizador do As Astrocientistas convidou $13$ mulheres a apresentarem seus trabalhos nas áreas de astrofísica, gravitação e cosmologia. A seleção das convidadas foi feita por todos os membros do comitê em conjunto, levando em consideração a abordagem de diferentes temas, diversidade regional e representatividade de mulheres pertencentes a outros grupos marginalizados. A grande dificuldade não foi encontrar pesquisadoras de excelência, mas sim selecionar um pequeno grupo em uma lista tão extensa. 

As convidadas da primeira edição do evento foram Angela Olinto (Universidade de Chicago), Beatriz B. Siffert (UFRJ), Carolina Loureiro Benone (UFPA), Dinalva Aires de Sales (IMEF - FURG), Elisa G.M. Ferreira (USP e Instituto Max Planck para Astrofísica), Leila Graef (UFF), Maria Elidaiana da Silva Pereira (Universidade de Hamburgo), Mariana Penna-Lima (UnB), Micol Benetti (Universidade de Nápoles Federico II), Miriani Pastoriza (UFRGS), Natalia Vale Asari (UFSC), Rita de Cássia dos Anjos (UFPR) e Vivian Miranda (Universidade do Arizona). Os títulos das palestras e uma breve introdução às carreiras acadêmicas das palestrantes convidadas podem ser encontrados no Apêndice \ref{app1}. Também convidamos a pesquisadora Érica de Mello Silva (UFTM), que integra o Grupo de Trabalho sobre Questões de Gênero da Sociedade Brasileira de Física desde 2019, para liderar uma discussão mais aprofundada sobre a participação de mulheres na ciência.

O fechamento do evento aconteceu no Dia Internacional de Meninas e Mulheres na Ciência, celebrado anualmente em 11 de fevereiro. Para a palestra de encerramento, tivemos a imensa honra de receber Andrea Ghez (Universidade da Califórnia), vencedora Prêmio Nobel de Física no ano de 2020. Além da oportunidade de ouvir sobre seu trabalho, vimos que a trajetória de uma mulher na ciência é desafiadora, ainda que se trate de uma futura Prêmio Nobel. O convite à Ghez, que é americana e apresentou em inglês, foi além da proposta inicial de homenagem à pesquisadoras brasileiras ou com vínculos com o Brasil. Dentre os $219$ laureados com o Nobel de Física até então, apenas $4$ são mulheres, correspondendo a aproximadamente 1,83\%. Por isso, julgamos que a participação da Ghez seria encorajadora, não só por ela ser uma mulher pesquisadora, mas porque sua contribuição científica é de extrema relevância para as astrociências. Ghez foi laureada pela descoberta de um objeto compacto supermassivo no centro de nossa galáxia \cite{Ghez.1998, Ghez.2008}, partilhando o prêmio com Roger Penrose e Reinhard Genzel.

O evento também contou com a apresentação de $11$ comunicações orais e $25$ pôsteres, cujos títulos e autores também podem ser encontrados no Apêndice \ref{app1}.

\section{O que aprendemos com o evento}
\label{pesquisa_evento}

Organizar o evento As Astrocientistas foi uma experiência enriquecedora e desafiadora em diversos aspectos. Além de todo o trabalho usual atribuído a comitês organizadores em geral, tivemos que nos atentar ao caráter social e político do encontro. Em um momento de tamanha intolerância e negacionismo em nosso país, quisemos criar um meio de discussão científica inclusivo e receptivo, assegurando o espaço e a participação ativa das minorias. Na prática encontramos diversos desafios a esse objetivo e, ao enfrentá-los, tivemos a chance de perceber com mais clareza algumas das dificuldades impostas às pessoas marginalizadas na ciência e, por fim, aprender com elas.

Um dos primeiros problemas que tivemos que endereçar foi a dificuldade de estabelecer o caráter científico do evento. Por mais que o site e as mídias sociais do encontro deixassem clara uma proposta acadêmica nas áreas de astrofísica, gravitação e cosmologia, muitos dos receptores da divulgação tomaram o evento única e estritamente como uma discussão sobre gênero. Esse foi, certamente, um tema abordado durante as discussões. Organizamos uma mesa redonda intitulada {Mulheres e Meninas na Ciência - Desafios e Oportunidades}, que contou com a participação das pesquisadoras Érica de Mello Silva, Dinalva Aires de Sales, Leila Graef e Micol Benetti, sem mencionar a ativa participação dos inscritos. Entretanto, o foco principal, desde o início, foi promover a interação e a discussão entre estudantes e pesquisadoras atuantes nas áreas mencionadas. Abrimos espaço para a discussão de temas científicos relevantes, com mesas redondas focadas no cenário atual das astrociências, com a participação de Caroline Benone, Natália Vale Asari e Vivian Miranda, e o futuro das pequisas em astrofísica  cosmologia e gravitação, cuja discussão foi liderada por Angela Olinto, Elisa Ferreira, Maria Elidaiana da Silva e Mariana Penna-Lima. 

Apesar da surpresa inicial que tivemos ao nos deparar com essa situação em que os objetivos do evento foram de certa forma mal interpretados, não é difícil relacioná-la ao constante distanciamento imposto entre o feminino e o científico. Da mesma forma que mulheres enfrentam obstáculos para serem identificadas como cientistas, um evento sobre pesquisas de mulheres enfrentou problemas para ser identificado como científico. Apesar de termos, por fim, alcançado o caráter almejado na primeira edição do encontro, hoje compreendemos que esforços extras nesse sentido são fundamentais.

\begin{figure}
	\begin{center}
		\includegraphics[width=7cm]{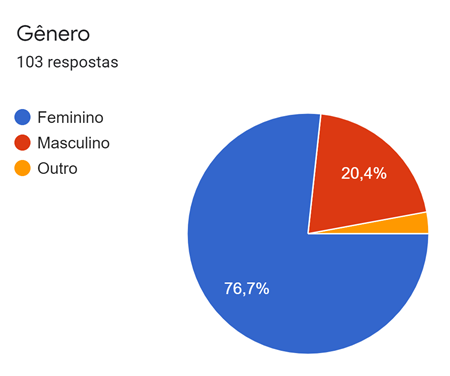}
	\end{center}
	\caption{\label{generoAA} Participação por gênero da primeira edição do evento {As Astrocientistas}.}
\end{figure}

A fim de aproveitar ao máximo a oportunidade apresentada com este tipo de evento, um contraespaço para dar voz às mulheres cientistas, coletamos algumas informações dos participantes através de uma pesquisa (opcional) disponível para participação durante esta primeira edição do evento As Astrocientistas. Os dados coletados sobre gênero confirmam a pouca participação masculina. Ao contrário da imensa maioria dos eventos acadêmicos na área da física, As Astrocientistas teve uma participação expressivamente maior de mulheres. Na Figura \ref{generoAA} apresentamos as porcentagens de participação por gênero.

\begin{figure}
	\begin{center}
		\includegraphics[width=7cm]{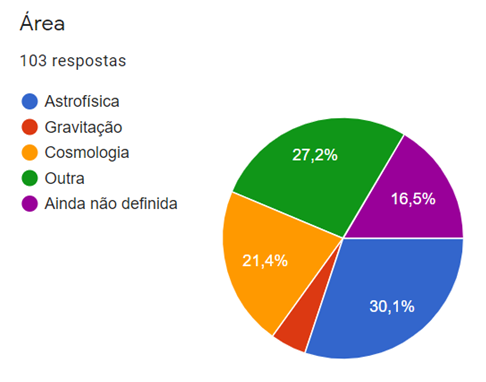}
	\end{center}
	\caption{\label{areasAA} Áreas de atuação dos inscritos na primeira edição do evento As Astrocientistas.}
\end{figure}

\begin{figure}
	\begin{center}
		\includegraphics[width=7cm]{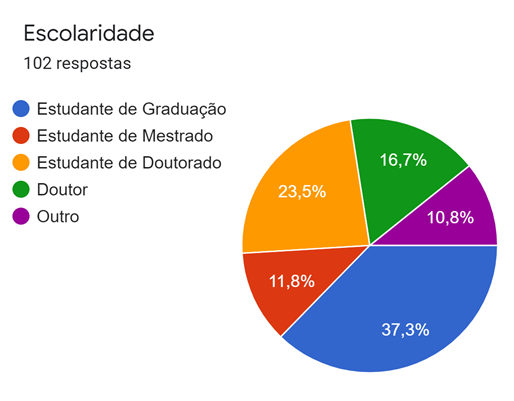}
	\end{center}
	\caption{\label{escolaridadeAA} Escolaridade dos inscritos na primeira edição do evento As Astrocientistas.}
\end{figure}

A pesquisa respondida pelos participantes durante o evento também nos forneceu uma informação extremamente relevante sobre as áreas de atuação dos inscritos. Na Figura \ref{areasAA} apresentamos as porcentagens de estudantes e pesquisadores atuantes nas diferentes áreas abordadas pelo encontro. A reduzida participação de pessoas da gravitação chama atenção, sobretudo pelo caráter teórico da área. Certamente precisamos endereçar essa questão nas edições futuras do evento, direcionando palestras a essa categoria que, até então, se encontra sub-representada. Com relação à escolaridade dos inscritos, foi possível abranger os diversos níveis acadêmicos de forma bastante equilibrada, como pode ser visto na Figura \ref{escolaridadeAA}.

Os dados também mostram uma concentração dos inscritos na região sudeste do Brasil, fenômeno já conhecido no cenário científico nacional. Apesar dos esforços do comitê organizador de incluir palestrantes e estudantes de diferentes regiões, ainda tivemos a maior parte dos inscritos sediada nos estados do Rio de Janeiro, Minas Gerais, São Paulo e Espírito Santo.

Mas talvez o maior aprendizado que tivemos foi com respeito à relevância de um evento como este. As conversas do comitê organizador durante o processo de organização foram bastante elucidantes, no sentido de que só quando pusemos a questão em palavras fomos capazes de compreender a dimensão do problema, apesar de vivenciarmos diariamente a desigualdade de gênero nas nossas áreas de pesquisa. Nós certamente falhamos em alguns aspectos, mas aprendemos que as vitórias fazem o esforço valer a pena.

\section{Conclusão: um convite para ação}

As Astrocientistas, I Encontro Brasileiro de Meninas e Mulheres na Astrofísica, Cosmologia e Gravitação, foi realizado de 08 a 11 de fevereiro de 2021 de forma virtual, de acordo com as normas de distanciamento social impostas para contenção da pandemia da COVID-19. O evento teve como objetivo homenagear as pesquisadoras brasileiras ou associadas ao Brasil num momento do nosso país onde a inversão de valores chegou a tal ponto que a ciência é negada e o preconceito celebrado. Durante a organização do evento, aprendemos que As Astrocientistas foi muito mais que uma homenagem; as palestras de altíssima qualidade e a ativa participação dos inscritos nos mostrou que este foi também um espaço de reflexão e colaboração para o combate à desigualdade de gênero dentro da academia.

Neste artigo, discutimos nosso aprendizado pessoal com a organização do evento As Astrocientistas e, para contextualizar essas questões, apresentamos de forma introdutória o cenário atual, global e brasileiro, identificando alguns fenômenos que se manifestam em diferentes estágios de formação acadêmica e dificultam o avanço de carreira de mulheres dentro da academia. Percebemos que, apesar da demanda crescente por estatísticas e análises comparáveis entre países sobre mulheres na ciência, os dados nacionais e seu uso na formulação de políticas costumam ser limitados. Mas que, felizmente, há um contínuo esforço para trazer luz para este problema tão debilitante para a ciência.

Dentro da academia, em particular em departamentos e institutos de física, há uma crença de que o cientista deve ser racional ao ponto de supressão de qualquer outro aspecto do consciente humano. Este estereótipo do cientista como um homem branco cuja personalidade é moldada em torno do seu entendimento de racionalidade é prejudicial para a ciência, pois diminui ou simplesmente apaga contribuições valiosas de pontos de vista diversos. Um resultado científico pode ser objetivo e removido de qualquer contexto humano, mas as práticas científicas não são. A perpetuação da ideia de que mulheres não são capazes de fazer ou não se interessam por ciência é prejudicial para o meio científico em mais de uma forma. Enquanto cientistas, devemos ter o entendimento do quanto a ciência é essencial para a sociedade e o futuro da nossa espécie. Mulheres correspondem a pouco mais da metade da população do nosso planeta, com capacidade para decidir, inclusive através do voto, a direção do nosso futuro. Desencorajar a participação e mesmo o interesse delas pela ciência se torna, então, contraintuitivo, visto que precisamos deste engajamento para continuar pautando nossas futuras decisões na ciência, que não é infalível, mas que é a aposta mais segura para deliberações políticas e econômicas.

As pesquisas internacionais evidenciam que, apesar de conseguirmos um progresso significativo, é possível que as desigualdades de gênero se repitam de outras formas, e por isso o debate constante do tema se faz cada vez mais necessário. Mudanças simples podem ajudar bastante a diminuir esta lacuna, como abrir oportunidades para a contratação de mulheres qualificadas, orientar uma mudança sugestiva no programa de mentores e analisar a possibilidade de patrocínio. Ou, no mínimo, encorajar a participação de grupos minoritários em processos seletivos. Pesquisas mais robustas também podem nos ajudar a compreender as implicações da relação entre hábitos culturais, gênero e pesquisas científicas. Na realidade, atuais avanços na discussão de igualdade de gênero na física já fornecem possíveis indicativos de como enfrentar o problema. Algumas ações são de fácil implementação, como por exemplo:
\begin{itemize}
	\item A valorização da diversidade dentro de uma equipe, em oposição ao enfoque em pesquisadores autossuficientes \cite{Freeman,Nielsen,Powell};
	\item Desenvolver mecanismos para evitar vieses contra grupos minoritários, o que pode ser feito através de treinamentos de conscientização.\footnote{Um exemplo de atuação deste mecanismo pode ser encontrado em \href{https://implicit.harvard.edu/implicit/}{https://implicit.harvard.edu/implicit/}.} Além da formação, também é importante observar a atuação dos vieses constantemente em encontros, seleções de comitês, e etc. \cite{Leru};
	\item Estar atento às microdesigualdades\footnote{Microdesigualdades é um termo cunhado por Sue Rosser para descrever as desvantagens sutis mas que desempenham um papel central em todos os estágios da carreira da mulher.} que ocorrem no dia a dia acadêmico e são extremamente prejudiciais para o desenvolvimento de relações saudáveis dentro de universidades e instituições de pesquisa \cite{Rosser};
	\item Levar em consideração perspectivas de gênero no desenvolvimento dos trabalhos de pesquisa. É necessário lembrar que ciência só acontece através de cientistas, então há um acoplamento entre objeto de estudo e sujeito, e por isso representatividade é fundamental;\footnote{Informações sobre como sexo/gênero podem ser características do que estudamos podem ser encontradas em \href{http://genderedinnovations.stanford.edu/}{http://genderedinnovations.stanford.edu/}.}
	\item A produção de contraespaços para dar suporte aos grupos que são marginalizados e discriminados.
\end{itemize}

Hábitos são difíceis de serem mudados e microagressões e vieses são um desafio para se medir, mas essas pequenas ações fazem muita diferença e podem ser aspectos-chave para alguém seguir ou não a carreira de cientista. Projetos como o evento As Astrocientistas, por exemplo, se encaixam especialmente neste último item citado. O encontro foi um contraespaço para dar voz a esta minoria de pesquisadoras que trabalham com astrofísica, cosmologia e gravitação. E o sucesso do evento foi completamente atribuído à participação e animação dos inscritos, mostrando mais uma vez o quanto podemos alcançar com um esforço direcionado.

\section{Agradecimentos}

Gostaríamos de agradecer primeiramente aos membros do comitê organizador, Emmanuel Frion e Rodrigo von Marttens, que abraçaram a causa e foram indispensáveis durante a organização do evento. Eles dividem os louros conosco, ativamente participando para a criação de um ambiente acadêmico diversificado. Agradecemos à Débora Menezes por nos dar permissão para apresentar os gráficos de sua pesquisa. Também estendemos os agradecimentos ao Prof. Nelson Pinto Neto, que muito gentilmente se prontificou a estabelecer nosso contato com a Prof$\,^{\text{a}}$ Angela Olinto, nos Estados Unidos.

PCMD agradece o financiamento Nº UMO-2018/30/Q/ST9/00795 do {National Science Centre}, Polônia. TM agradece ao Prof. Tomas Brage por uma conversa estimulante e referências após uma sessão especial de diversidade apresentada no evento {Physics Days 2021}, que ocorreu virtualmente na Universidade de Jyväskylä.

Por fim, nosso muito obrigada a todos que participaram do evento As Astrocientistas, professoras, jovens pesquisadoras e a todos os inscritos. Vocês foram a razão do sucesso do evento.

%
%

\appendix
\section{Palestras, comunicações orais e pôsteres}\label{app1}

Importantes contribuições foram dadas por pesquisadoras, pós-graduandos e estudantes em diversos níveis acadêmicos. A seguir apresentamos os títulos e autores das palestras, comunicações orais e pôsteres que participaram do evento As Astrocientistas em fevereiro de 2021.

\subsection{Palestras}
\begin{itemize}
	\item Multi-mensageiros cósmicos e suas mensagens na física de astropartículas - Angela Olinto (Universidade de Chicago).
	\bigskip
	\\
	Angela Olinto é doutora em física pelo Instituto de Tecnologia de Massachusetts (MIT) e atuou em várias posições de destaque no cenário internacional, sendo inclusive diretora do Departamento de Astronomia e Astrofísica da Universidade of Chicago. Também foi vencedora de diversos prêmios e, mais recentemente, tornou-se membro da Academia Americana de Artes e Ciências. Seu trabalho tem como foco raios cósmicos de altas energias, assinatura indireta de partículas de matéria escura, efeitos cosmológicos de campos magnéticos, inflação natural e a estrutura interna de estrelas de nêutron, dentre vários outros interesses. Ela está à frente do projeto POEMMA, sigla em ingês para Sonda Astrofísica Multi-Mensageira \cite{Olinto.2021}.
	
	
	\item Astrobiologia: descobrindo novos planetas - Beatriz B. Siffert (UFRJ).
	\bigskip
	\\
	Beatriz Siffert é doutora em física pela Universidade Federal do Rio de Janeiro. Tem experiência nas áreas de cosmologia e astrofísica, tendo trabalhado com detecção indireta de matéria escura, modelos cosmológicos com ricochete, e, atualmente, com supernovas do tipo Ia \cite{Coelho_2014} e astrobiologia. É professora adjunta de física no Campus Duque de Caxias da UFRJ.
	
	\item Análogos gravitacionais: Histórico, modelos e perspectivas - Carolina Loureira Benone (UFPA).
	\bigskip
	\\
	Carolina Loureiro Benone é doutora em física pela Universidade Federal do Pará, com período sanduíche na Universidade de Aveiro. Atualmente é professora adjunta da Universidade Federal do Pará - Campus Salinópolis. Seu trabalho tem ênfase em relatividade geral, trabalhando com modelos análogos gravitacionais \cite{Benone.2020}.
	
	\item AstroBioGeoQuímica: estudo de moléculas orgânicas em galáxias com buracos negros supermassivos e suas consequências para a vida na Terra - Dinalva Aires de Salles (IFME/FURG).
	\bigskip
	\\
	Dinalva Aires de Sales é doutora em física pela Universidade Federal do Rio Grande do Sul. Tem experiência na área de astrofísica, atuando principalmente nos seguintes temas: núcleo ativo de galáxia, galáxia em interação, moléculas complexas, espectroscopia e imagiamento \cite{Sales_2010}. Atualmente é professora adjunta e coordenadora do Programa de Pós-Graduação em Física do Instituto de Matemática, Estatística e Física da Universidade Federal do Rio Grande.
	
	\item A natureza da matéria escura e  a matéria escura ultra-leve - Elisa G.M. Ferreira (Instituto Max Planck para Astrofísica).
	\bigskip
	\\
	Elisa Ferreira é doutora pela Universidade McGill e atualmente é professora no Instituto de Física da Universidade de São Paulo, sendo também associada ao Instituto Max Planck para Astrofísica. Sua área principal de pesquisa é a cosmologia, com foco no estudo da natureza da matéria escura e da energia escura e em como esses modelos podem ser testados com as observações atuais e futuras \cite{ElisaF}. Ela também é membro de colaborações internacionais como o {Prime Focus Spectrograph} e o Telescópio BINGO.
	
	\item Conectando o universo primordial e recente através das ondas gravitacionais e eletromagnéticas - Leila Graef (UFF).
	\bigskip
	\\
	Leila  Graef é doutora em física pela Universidade de São Paulo, com período sanduíche na Univesidade McGill. Atualmente é professora adjunta no Instituto de Física da Universidade Federal Fluminense e trabalha na área de cosmologia, atuando principalmente nos seguintes temas: universo primordial, perturbações cosmológicas, radiação cósmica de fundo, energia escura e gravitação quântica \cite{Graef.2020}.
	
	\item Pesando gigantes no céu: Cosmologia com aglomerados de galáxias - Maria Elidaiana da Silva Pereira (University of Michigan).
	\bigskip
	\\
	Maria Elidaiana da Silva Pereira é doutora em física pelo Centro Brasileiro de Pesquisas Físicas. Atualmente faz pós-doutorado na Universidade de Hamburgo. Realiza pesquisas na área de cosmologia observacional, estudando o efeito de lenteamento gravitacional, aglomerados de galáxias e ondas gravitacionais \cite{Pereira.2017}.
	
	\item Cosmologia com aglomerados de galáxias - Mariana Penna-Lima (UnB).
	\bigskip
	\\
	Mariana Penna-Lima é doutora em física pelo Centro Brasileiro de Pesquisas Físicas. Tem experiência nas áreas de cosmologia e gravitação \cite{Penna-Lima.2021}. Atualmente é Professora Adjunta do Instituto de Física da Universidade de Brasília. Além disso, é membro das colaborações internacionais LSST/DESC e JPAS. 
	
	\item Cosmologia hoje e objetivos para as próximas décadas - Micol Benetti (University of Naples Federico II).
	\bigskip
	\\
	Micol Benetti é doutora em astrofísica relativística pela Universidade de Roma, La Sapienza. Fez pós-doutorado no Observatório Nacional e atualmente faz pós-doutorado na Universidade de Nápoles Federico II. Atua na área de cosmologia, tendo trabalhado com modelos de universo primordial, extensões do Modelo Cosmológico Padrão e análise de dados a fim de vincular propriedades do universo primordial e tardio e de física fundamental através de observações cosmológicas \cite{Benetti_2021}.
	
	\item Interações de galáxias: efeitos sobre cinemática do gás  e a formação de estrelas - Miriani Pastoriza (UFRGS).
	\bigskip
	\\
	Miriani Pastoriza é doutora em astronomia pela Universidade Nacional de Córdoba. Atualmente é professora colaboradora do Departamento de Astronomia do Instituto de Física da Universidade Federal do Rio Grande do Sul, onde também é professora emérita. Tem experiência na área de astronomia, com ênfase em astrofísica extragaláctica, atuando principalmente nos seguintes temas: galáxias, formação estelar, núcleos ativos e interação de galáxias. Além disso, é membro titular da Academia Brasileira de Ciências. Uma de suas grandes contribuições científicas foi a descoberta e caracterização das galáxias Sérsic-Pastoriza, juntamente com o astrônomo José Luis Sérsic \cite{S_rsic_1965}.
	
	\item O ciclo de vida das estrelas dá o compasso do enriquecimento químico de galáxias - Natalia Vale Asari (UFSC).
	\bigskip
	\\
	Natalia Vale Asari possui duplo doutorado em cotutela: doutorado em física, área de concentração astrofísica, pela Universidade Federal de Santa Catarina e doutorado em Astronomia e Astrofísica pelo Observatório de Paris. Tem experiência na área de astrofísica, com ênfase em astrofísica extragaláctica, atuando principalmente nos seguintes temas: evolução de galáxias, atenuação por poeira, espectroscopia-3D, síntese espectral e núcleos ativos de galáxias. Atualmente é Professora Adjunta na Universidade Federal de Santa Catarina \cite{ValeAsari.2017}.
	
	\item Galáxias starburst e rádio-galáxias como fontes de raios cósmicos de altíssimas energias - Rita de Cássia dos Anjos (UFPR).
	\bigskip
	\\
	Rita de Cássia dos Anjos é doutora em física pela Universidade de São Paulo, São Carlos. Atualmente é professora da Universidade Federal do Paraná no Setor Palotina e trabalha com raios cósmicos de energias acima de 1EeV (Observatório Pierre Auger) e energias entre 10GeV e 100TeV (Cherenkov Telescope Array - CTA). Seu trabalho tem ênfase no estudo da propagação de raios cósmicos e interações de partículas e raios gama e no estudo da Equação de Schrödinger (Fokker Planck) \cite{dos_Anjos_2021}. É membro do Observatório de Raios Cósmicos Pierre Auger e membro do Observatório { Cherenkov Telescope Array} (CTA). 
	
	\item O Universo conectado: relacionando o universo inicial, intermediário e tardio com dados cosmológicos - Vivian Miranda (Universidade of Arizona).
	\bigskip
	\\
	Vivian Miranda é doutora pela Universidade de Chicago e atualmente faz pós-doutorado no Observatório Steward, da Universidade do Arizona. Além disso, colabora com o DESC-LSST. Em 2019, conquistou o Prêmio {Leona Woods Distinguished Postdoctoral Lectureship}. Ela investiga como as teorias da inflação e da energia escura podem ser testadas com os dados observacionais \cite{Miranda.2021}.
\end{itemize}

\subsection{Comunicações orais}
\begin{enumerate}
	\item Recuperando atmosferas de exoplanetas utilizando machine learning - Aline Novais (UFRJ);
	
	\item Desvendando o meio interestelar de galáxias empoeiradas: uma análise astroquímica - Yanna Martins-Franco (OV/UFRJ);
	
	\item Medidas de abundâncias a partir do método direto para galáxias com formação estelar ativa - Katia Slodkowski Clerici (UFSC);
	
	\item Modelo cosmológico intrinsecamente simétrico na presença de fluidos dissipativos - Grasiele Batista dos Santos (UNIFEI);
	
	\item Singularidades futuras repentinas e partículas - Olesya Galkina (UFES);
	
	\item Perspectivas para a física de ultralargas escalas: inflação e efeitos relativísticos - Caroline Macedo Guandalin (IFUSP);
	
	\item Primeiros passos na obtenção de parâmetros cosmológicos utilizando matrizes de covariância cosmológicas sem ruído - Natalí Soler Matubaro de Santi (USP);
	
	\item Skewness como um teste para a energia escura - Raquel Fazolo (PPGCosmo);
	
	\item Nova parametrização para a energia escura baseada no gás de Chaplygin generalizado - Dinorah Barbosa da Fonseca Texeira (ON);
	
	\item Relatividade especial invariante sob o grupo de de Sitter e as curvas de rotação de galáxias - Adriana Victoria (Universidad Sergio Arboleda);
	
	\item Trajetória e pesquisas de Eliade Lima - Lucas Bicalho (UESB);
\end{enumerate}

\subsection{Pôsteres}
\begin{enumerate}
	\item Análise elementar do Meteorito
	condrito carbonáceo Águas Zarcas - Aisha Alana Persaud Leitch (UFRJ);
	
	\item Tópicos da astronomia que contestam a teoria da Terra plana - Alice Taís Dummel Weide (EEEM Guararapes);
	
	\item Ensino e divulgação de astrobiologia no ensino médio - Angela Ferreira Portella (UFRJ);
	
	\item Caracterização mineralógica do meteorito brasileiro Serra Pelada e suas implicações para a história geológica do asteroide (4) Vesta - Bruna Mayato Rodrigues (Museu Nacional/UFRJ);
	
	\item Sobre quasares, a formação de estruturas no Universo e tudo o mais - Carolina Queiroz de Abreu Silva (USP);
	
	\item Projeto meninas na ciência incentiva a formação da identidade do grupo ``Meninas do Guara'' - Cristine Inês Brauwers (EEEM Guararapes);
	
	\item Campos magnéticos e seus efeitos na
	propagação de partículas - Débora Beatriz Götz (UFPR);
	
	\item Propriedades do contínuo e do gás emissor
	de linhas largas em AGNs emissores de FeII - Denimara Dias dos Santos (INPE);
	
	\item Terraformação de Marte - Gabriela Medeiros de Carvalho (IFCE);
	
	\item Base Lunar $-$ Divulgação de astronomia através de ilustrações $-$ Ingrid dos Santos Beloto (IAG - USP);
	
	\item Propriedades Fotométricas de
	Grupos e Aglomerados de Galáxias - Kethelin Parra Ramos (USP);
	
	\item Um estudo comparativo do cometa C/1977 R1 (Kohler): Analisando fosseis do Sistema Solar - Loreany Ferreira de Araújo (USP);
	
	\item Fontes de raios cósmicos de altíssimas energias - Luana Natalie Padilha (UTFPR);
	
	\item Livro sobre buracos negros: de aluna para alunos - Luísa Machado Saldanha;
	
	\item Meninas do Guara e a experiência das sessões de observação abertas à comunidade - Maiara Graff (EEEM Guararapes);
	
	\item Anisotropia de raios cósmicos de Altíssimas Energias - Maria Clara Dari Gomes (UFPR);
	
	\item Identifying protoclusters in distant universe - Mariana Rubet (UFRJ);
	
	\item Estudo da formação de exoplanetas terrestres em regiões de ZH de sistemas binários - Mayra Meirelles Marques (UFRJ);
	
	\item Magnetic effects of electrical discharges on Mars - Melissa de Andrade Nunes (IAG - USP);
	
	\item Em busca de sinais de crescimento de buracos negros supermassivos em galáxias ultracompactas brilhantes no ultravioleta usando espectroscopia no infravermelho médio - Rayssa Guimarães Silva (Observatório do Valongo - UFRJ);
	
	\item Um breve estudo sobre a estrutura das estrelas compactas - Tatiane Corrêa (UERJ); 
	
	\item Análise química e mineralógica dos condritos carbonáceos Allende e Murchison: implicações no surgimento da vida no Universo - Tatiane Peters Donato (Museu Nacional - UFRJ);
	
	\item Relatividade especial de deSitter: implicações para a massa de aglomerados de galáxias - Thais Campos Santiago (IFT-UNESP);
	
	\item Rotating neutron stars and modified gravity: Scalar charges and pulsar-timing observables in the presence of nonminimally coupled
	scalar fields - Tulio Ottoni Ferreira da Costa (PPGCosmo);
	
	\item Analema: descrição da trajetória aparente do Sol durante o período de translação da Terra - Yalle Carolina Ferreira da Silva (UESB);
\end{enumerate}

\bibliography{main}

\end{document}